# Electrically controlled water permeation through graphene oxide membranes

K.-G. Zhou[1,2*], K. S. Vasu[1,2*], C. T. Cherian[1,2], M. Neek-Amal[3,4], J. C. Zhang[5], H. Ghorbanfekr-Kalashami[4], K. Huang[1,2], O. P. Marshall[6], V. G. Kravets[6], J. Abraham[1,2], Y. Su[1,2], A. N. Grigorenko[6], A. Pratt[5], A. K. Geim[6], F. M. Peeters[4], K. S. Novoselov[6], R. R. Nair[1,2*]

[1]National Graphene Institute, University of Manchester, Manchester, M13 9PL, UK.

[2]School of Chemical Engineering and Analytical Science, University of Manchester, Manchester, M13 9PL, UK.

[3]Department of Physics, Shahid Rajaee Teacher Training University, 16875-163, Lavizan, Tehran, Iran.

[4]Department of Physics, University of Antwerpen, Groenenborgerlaan 171, B-2020 Antwerpen, Belgium.

[5]Department of Physics, University of York, York, YO10 5DD, UK.

[6]School of Physics and Astronomy, University of Manchester, Manchester M13 9PL, UK.

*kai-ge.zhou@manchester.ac.uk, vasusiddeswara.kalangi@manchester.ac.uk, rahul@manchester.ac.uk

**Developing 'smart' membranes that allow precise and reversible control of molecular permeation using external stimuli would be of intense interest for many areas of science: from physics and chemistry to life-sciences[1-10]. In particular, electrical control of water permeation through membranes is a long-sought objective and is of crucial importance for healthcare and related areas. Currently, such adjustable membranes are limited to the modulation of wetting of the membranes[5] and controlled ion transport[1], but not the controlled mass flow of water. Despite intensive theoretical work[6-9,11-14] yielding conflicting results, the experimental realisation of electrically controlled water permeation has not yet been achieved. Here we report electrically controlled water permeation through micrometre-thick graphene oxide (GO) membranes. By controllable electric breakdown—conductive filaments are created in the GO membrane. The electric field concentrated around such current carrying filaments leads to controllable ionisation of water molecules in graphene capillaries, allowing precise control of water permeation: from ultrafast permeation to complete blocking. Our work opens up an avenue for developing smart membrane technologies and can revolutionize the field of artificial biological systems, tissue engineering and filtration.**

Controlled transportation of water molecules through membranes and capillaries is a ubiquitous natural phenomenon that is absolutely vital to all kinds of living organisms[2,15,16]. Thus, an ability to create nanostructured membranes where water transport is controlled by an external parameter is of crucial importance for life science and healthcare. Such membranes would allow creation of artificial biological membranes and give a significant boost to research in tissue engineering. Of particular interest are membranes where water permeation could be controlled by an electric field, allowing fast and reliable response and easy integration into more complex systems. Previous attempts to control water permeation through membranes, mainly polymeric, were concentrated on modulating the membrane structure and physicochemical properties of the membrane surface by varying the pH, temperature and ionic strength[3,4,17]. Further, electrically controlled water permeation has been discussed in numerous theoretical and molecular dynamic (MD) simulation studies[6-12], which, however, often yield conflicting results ranging from freezing of water molecules to melting of ice under an electric field[12-14].

The latest advances in the fabrication of artificial channels and membranes using two-dimensional (2D) materials[18-23] have led to the intensive study of nanoscale and sub-nm-scale water and ion transport. In particular, GO membranes containing 2D graphene capillaries exhibit ultrafast permeation of water[18] and unique molecular sieving properties[19,23] with the potential for industrial-scale production. In this report, we study electrical effects on water permeation through GO membranes, and demonstrate that water flux through such membranes can be controlled over a wide range by an applied electric field.

Our devices are essentially GO membranes with metal electrodes on both sides. Such metal/GO/metal sandwich structures were fabricated by depositing a thin (≈ 10 nm) gold (Au) film on top of the GO membrane prepared on a porous silver (Ag) substrate (Methods and supplementary Fig. 1 and 2). Such a thin layer of gold is sufficiently porous and does not change the permeation properties of the membranes. Figures 1a, b show a schematic and an optical photograph of our membrane device. These sandwich structures were glued onto a plastic disc with a circular aperture (~10 mm diameter) and then used to seal a water-filled steel container to expose the device to water vapour and allow the measurement of water permeation using gravimetric analysis[18,20]. A DC bias voltage was applied across the membrane using a Keithley 2410 sourcemeter (Fig. 1a) with the current compliance (20 mA for the sample in Fig. 1) set to prevent uncontrollable breakdown.

In order to increase the electric field applied up to the values sufficient for water dissociation, thin conductive filaments were introduced in our membranes by controllable electric breakdown. It is known that in the presence of moisture, the formation of permanent conducting paths (often carbon) across the surface of the insulator occurs when a large electric field is applied[24]. We used this phenomenon for the formation of conducting filaments inside GO membranes (supplementary section 2 and supplementary Figs. 3-7). Fig. 1b shows the *I-V* characteristics of the device during the filament formation. Up to a critical voltage, $V_c$, the current does not change appreciably. However, at $V_c$ (~2 V for the sample in Fig. 1 and varies by 25% from sample to sample; four samples were studied) a partial electrical breakdown occurred, evident through a sudden current increase up to the compliance level. This breakdown state is stable and is characterised by low transversal resistance.

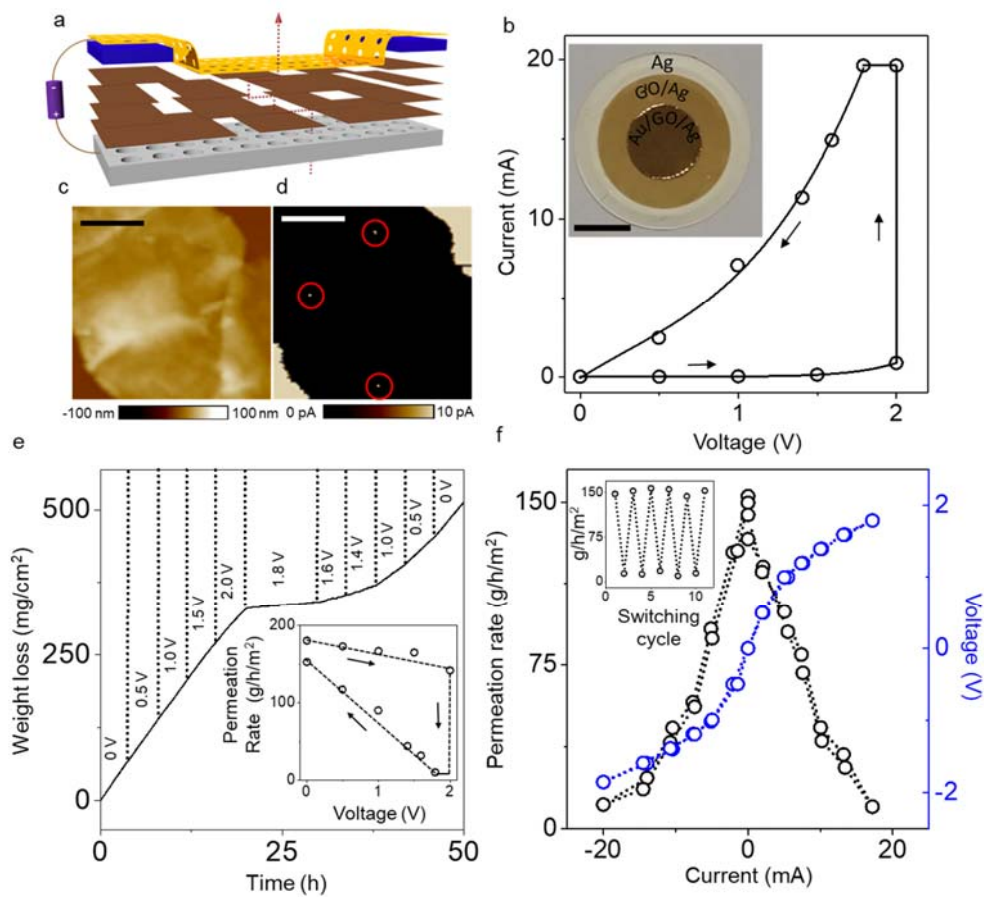

**Figure 1| Electrically controlled water permeation through a graphene oxide membrane.** (**a**) Schematic showing a GO membrane (deposited on a porous silver substrate) with voltage applied. The top yellow layer represents a porous gold electrode, the blue layer is a plastic mask used to avoid electric shorting between gold and silver electrodes, and the brown layers represent GO sheets. The dotted line shows a possible pathway for water permeation. (**b**) *I-V* characteristics during the first voltage sweep showing a sudden increase in current suggesting partial electrical breakdown of the GO membrane and conducting filament formation. The solid line is a guide to the eye and arrows indicate the direction of the voltage sweep. Inset: Photograph of a GO membrane showing the central sandwiched (Au/GO/Ag) region, the GO on supporting silver and the bare silver substrate. The plastic mask outside the gold layer is removed for clarity. The outer area of the central sandwiched region was masked to block water permeation through this region (supplementary Fig. 1 and 2). Scale bar: 5 mm. (**c**) Topography and the corresponding PF TUNA current image (**d**) of a conducting GO membrane exfoliated on a gold thin film coated Si substrate. The conducting filaments (size ranging from 20 to 45 nm) formed in the GO membrane are marked by red circles. Scale bars: 1 μm. (**e**) Weight loss for a water-filled container sealed with a 1 μm thick GO membrane (7 mm diameter) at different voltages applied across the membrane during the filament formation process. Inset: Water permeation through the GO membrane as a function of applied voltage. Arrows indicate the direction of voltage sweep. (**f**) Variation of water permeation rate as a function of the current across the GO membrane after the membrane became electrically conducting due to filament formation, and the corresponding *I-V* characteristics (color-coded axis). One complete voltage sweep for both positive and negative polarity is plotted. Inset: Continuous switching between 0 and 1.8 V showing the stability of permeation control. All weight loss measurements were performed inside a dry chamber with 10% relative humidity.

After the controllable breakdown, the *I-V* characteristics exhibit nearly ohmic behaviour and suggest the appearance of permanent electrically conductive channels (Fig. 1b). However, the in-plane *I-V* measurements do not show any significant change in conductivity when compared to the pristine samples (supplementary Fig. 4). Unlike in-plane conductivity, the out-of-plane conductivity was found to be stable and insensitive to the humidity of the environment (supplementary Fig. 4). This confirms the formation of conducting filaments (e.g., carbon filaments) between the electrodes (out-of-plane) that are not connected in the plane of the membrane. To characterise the filaments in the GO membrane we performed peak force tunnelling AFM (PF TUNA) and Raman spectroscopy (supplementary section 2 and supplementary Fig. 5 and 6). Fig. 1d shows one of the PF TUNA current images, displaying the presence of conductive filaments of diameter < 50 nm. The estimated filament density from the PF TUNA and Raman is ~$10^7$ per $cm^2$ (supplementary section 2). Note that no significant changes were observed in the chemical stoichiometry of the GO membrane (as measured by X-ray photoelectron spectroscopy (XPS)), except for a small increase in the C/O ratio (3.2 to 3.6) on the membrane surface closest to the positive electrode. The C/O ratio remains the same as for the pristine sample for all other membrane surfaces probed (supplementary section 3 and supplementary Fig. 8).

To probe the influence of electric field on the water permeation through the pristine GO membrane, we have monitored it during the filament formation process. The applied voltage was increased stepwise to enable both current and water permeation to be monitored as a function of time. At each voltage step, measurements were carried out for a minimum period of four hours. Fig. 1e shows the weight loss of the sealed container and the corresponding water permeation rate during the filament formation process. No appreciable change in water permeation was found up to $V_c$ (Fig. 1e). At $V_c$, a sudden decrease (15 times) in water permeation was observed after the partial breakdown of GO membrane. Hereafter, the water permeation through the membranes with the low transverse resistance has shown a strong dependence on the applied voltage, decreasing with increased voltage. At zero voltage, the water permeation practically recovered (~85%) to the initial value of the pristine sample.

The stable out-of-plane electrical conductivity of the membrane and the electrical control of water permeation are more evident from the continuous forward and backward voltage sweeps performed after the filament formation (Fig. 1f). The electrically controlled water permeation is found to be independent of the polarity of the applied voltage and is completely reversible even after multiple voltage cycles (Fig. 1f inset). Continuously switching the DC bias voltage between 0 and 1.8 V also demonstrates the membrane's durability and also the ability to controllably switch water permeation between 'on' and 'off' states in a precise manner. The electrical control of water permeation was further confirmed by mass spectrometry (supplementary section 4) which showed that no gas release occurred during the experiments (supplementary Fig. 9).

To understand the influence of voltage and current on the permeation, we have conducted two sets of additional experiments. First, we performed electrically controlled water permeation experiments on a GO membrane device with the same thickness but a four times smaller permeation area (supplementary section 5). To achieve the same level of water blockage as in the larger membrane, the smaller membrane device only requires a quarter of the current compared to the large device (supplementary Fig. 10). This indicates that current density through the membrane is the crucial factor in controlling water permeation. Second, we performed water permeation experiments using membranes with the same permeation area but a different thickness (1 and 5 μm). The decrease in water permeation rate was found to be nearly identical for the same magnitude of electric current passing through 1 μm and 5 μm thick membranes (supplementary Fig. 10). Importantly, the water permeation rate followed the variation in current rather than applied voltage (supplementary section 5). Both of these observations suggest that water transport in these experiments is predominantly controlled by the value of the current through conductive filaments rather than the voltage applied.

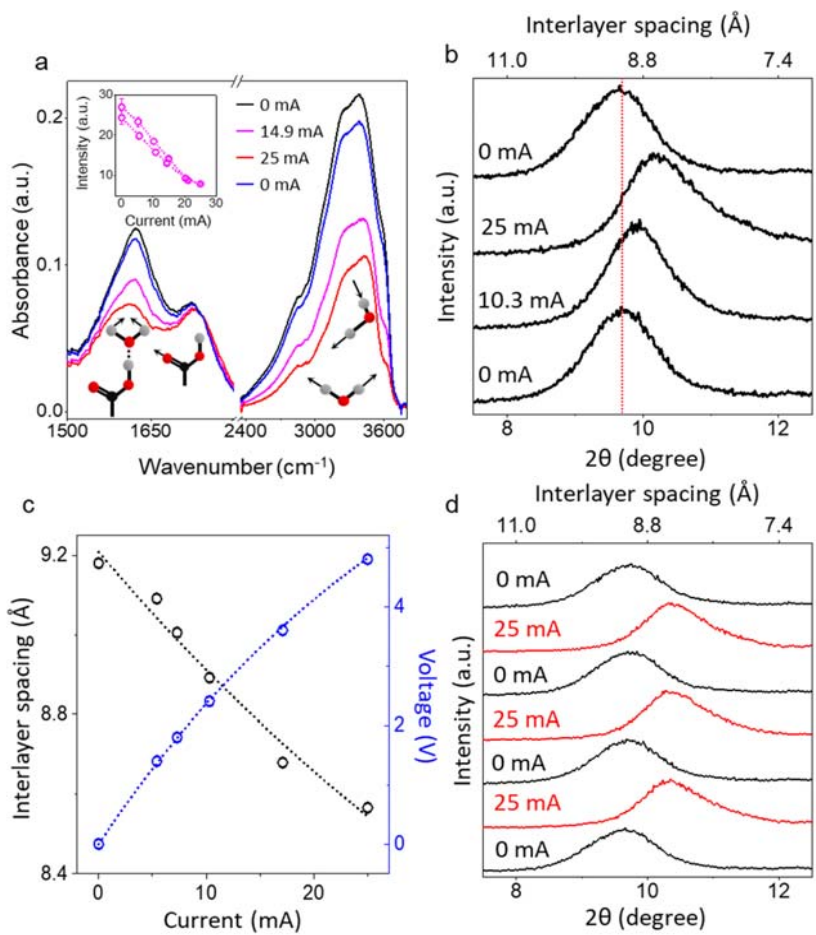

**Figure 2| *In-situ* FTIR and X-ray measurements.** (**a**) In-situ FTIR spectrum acquired from a GO membrane in which the current was cycled from 0 mA to 25 mA and then back to 0 mA. Inset: Peak intensity (area under the curve) as a function of the current for the peak at ~3500 cm$^{-1}$ for forward and backward sweeps. The ball-and-stick models below the data illustrate the vibrational modes responsible for the different IR peaks. Red – Oxygen, Black – Carbon, and Grey – Hydrogen. (**b**) X-ray diffraction for different current levels across the 10 μm thick GO membrane (membrane diameter ~10 mm). (**c**) Change in the interlayer spacing as a function of current and the corresponding *I-V* behavior of the membrane (color-coded axis). (**d**) Reversible switching of the (001) X-ray peak with the current.

The influence of the current on the water transport can be either through Joule heating or possible electrochemical changes. We have measured the membrane temperature as a function of the electric current across the membrane (supplementary section 6) and found no significant variations (supplementary Fig. 11). In order to test the electrochemical mechanism, we measured the water infrared (IR) modes from the GO membrane for a varying electric current across the membrane (Methods and Figure 2a). The pristine sample shows three main characteristic IR peaks at ~1620, 1737 and 3500 cm$^{-1}$.[25-27] The band at ~1620 cm$^{-1}$ is assigned to the deformation vibration of adsorbed water molecules and the band at ~1737 cm$^{-1}$ is due to the carbonyl (C=O) stretching mode of the carboxylic group itself[26,27]. The broad band between 3000 to 3500 cm$^{-1}$ is due to the O-H stretching mode in both the GO sheets and the interspersed water molecules[25]. When an electric current is switched on, all the band intensities corresponding to water molecules decrease whereas the band intensity relating to carbonyl groups remains constant. After the current was brought to zero, the bands associated with water molecules fully recover to their initial intensities (Fig. 2a inset).

To test if the decrease in the IR water signal is associated with the reduced amount of water, we performed an X-ray diffraction (XRD) study of the interlayer distance[23] in our devices as a function of the electric current (Methods). Fig. 2b shows the *in-situ* changes of the (001) reflection as a function of the current across the membrane with the (001) peak clearly shifting to higher 2θ values at elevated currents. The interlayer spacing, $d$, is estimated from the XRD analysis and plotted as a function of electric current in Fig. 2c. We found that a value of $d \approx 9.2$ Å at zero current decreases to $\approx 8.5$ Å as the current across the membrane increases from zero to 25 mA. These electric current induced changes in $d$ were also found to be reversible (Fig. 2d). Note that such changes in the interlayer distance were not present during the first voltage sweep up to $V_c$, before the filament formation. However, the observed small change in $d$ (0.7 Å) is not expected to affect the water permeation largely due to the slip-enhanced water permeation through graphene capillaries in the membrane[20,23]. Also, such a small decrease in the interlayer distance cannot explain the observed reduction in the IR peak intensity by nearly 50%.

Based on these observations, we attribute the electrically controlled water permeation to current-mediated ionisation of water molecules. It is known that a current-carrying conductor produces an electric field around it[28,29] (supplementary section 7). The exact value of the field depends on the parameter of the set-up, but for a coaxial arrangement (when current flows in one direction through the inner conductor of radius $a$ and in the other direction through a coaxial conductor of radius $b$— in our case the characteristic size of the sample), the radial component of the field is given by

$$E_r = \frac{Jz}{\sigma r} \frac{1}{\ln(a/b)}$$

where $J$ is the current density, $z$ varies from 0 to $L$ where $L$ is the length of the wire (thickness of the membrane), $r$ is the radial distance from the centre of the wire, and $\sigma$ is the electrical conductivity of the wire[28,29] (supplementary section 7). In our case this formula reduces to $E_r = (V/r)\cdot(1/\ln(a/b))$, and it is obvious that close to the surface of the filaments ($r$ is of the order of few tens nm) the electric field can be as high as $\sim 10^7$ V/m (supplementary Fig. 12). Such large electric fields could dissociate water molecules to produce hydronium and hydroxyl ions (supplementary section 8), with the effect becoming more pronounced at higher currents. Drift of such charged ions in the channels can suppress the water flow (supplementary section 8).

Our MD simulations (supplementary section 8) further confirm that the water permeation rate in graphene capillaries decreases with increasing concentration of hydronium and hydroxyl ions, supporting the observed electrical control of water permeation (supplementary Fig. 13). The proposed model of electric field enhanced dissociation of water is also consistent with the observed reversible chemical changes in the interlayer water molecules (IR peak intensities in Fig. 2a) and the changes in the interlayer spacing (Fig. 2b) with current. This could be attributed to the decrease in the volume of interlayer water due to ionisation[30].

In summary, the unforeseen electrical control of water permeation through GO membranes has profound significance in membrane-based separation technologies, nanofluidics and biomedical applications where the precise delivery of water molecules is crucial. Further experimental and theoretical efforts are needed for a more detailed and thorough understanding of the exact mechanism through which water permeation is controlled by electrical current. However, the research reported here is an important step in the understanding of water in nanoscale capillaries and its precise and reversible control via a simple means without involving complicated chemical modifications.

**Methods**

**Fabrication of metal-GO-metal sandwich membranes and electrical measurements:** GO aqueous dispersions (flake size of ≈ 10 μm) were prepared, as reported earlier[18,23], from the exfoliation of graphite oxide powder (BGT Materials Limited) in water using bath sonication. Different steps in the fabrication process of the metal-GO-metal sandwich membrane are shown in supplementary Fig. 1. First, a Sterlitech porous silver metal membrane (0.2 μm pore size and 13 mm diameter), which also acts as the bottom electrode, was used in a standard vacuum filtration set-up to prepare GO membranes for the permeation experiments. These membranes were glued (using Stycast 1266 epoxy resin) onto the polyethylene terephthalate (PET) films (step 1 in supplementary Fig. 1) with a circular aperture (diameter of ~1 cm) before evaporating a thin gold film (~10 nm) as the top electrode (step 2 in supplementary Fig. 1 and supplementary Fig. 2) to allow an electric potential to be applied across the membrane. Here, the PET film prevents electrical shorting between the top and bottom electrodes in the metal-GO-metal sandwich membrane structures. Finally, these structures were glued onto another plastic disc, as shown in supplementary Fig. 1 (step 3), which provides mechanical support for sealing the sample to a stainless steel water container used in pervaporation experiments[18]. Thin copper wires extending from both electrodes were connected to a Keithley 2410 sourcemeter (supplementary Fig. 1) to provide a DC voltage across the membranes allowing electrical effects on water permeation to be probed. An additional Keithley 2182A Nanovoltmeter was also connected across the membrane to measure the potential drop.

**Permeation experiments:** We employed a previously reported gravimetric method[18] to understand the electrical effects on water permeation through GO membranes. The plastic disc containing a metal-GO-metal sandwich membrane was fixed to a stainless steel water container using two rubber O-rings to ensure an air-tight seal. The water permeation rate was measured in terms of weight loss of the water container using a computer-controlled precision balance (Mettler Toledo; accuracy 0.1 mg). All the gravimetric experiments were carried out in a chamber with a controlled relative humidity of 10%. To probe electrical effects on water permeation, a DC voltage was applied across the membranes as shown in supplementary Fig. 1.

**In-situ IR measurements:** We have used *in-situ* IR absorption spectroscopy[25] to monitor any chemical changes in the GO membranes during voltage cycling. The IR measurements were performed in transmission geometry (typically 512 scans per loop) by employing VERTEX 80, Bruker FT-IR spectrometer and HYPERION Microscope, using a MCT (mercury cadmium telluride) liquid $N_2$ cooled detector with a mirror optical velocity of 0.6329 cm/s at a resolution of 4 $cm^{-1}$. To enable an electric potential to be applied across the membrane during the IR measurements, we deposited thin (10 nm) gold electrodes on both sides of the freestanding GO membrane. During the IR measurement, the whole system was continuously purged with a dry $N_2$ stream to remove water in the atmosphere. To match the IR experimental conditions to those of the permeation experiments, we used a water reservoir (a drop of water) at the edge of the GO membranes (away from the IR spot) to hydrate the membrane during measurements. Without this reservoir, IR spectra from the samples resemble that of a dry GO membrane (lower OH vibration peak).

**In-situ XRD measurements:** For probing the changes in the interlayer distance of GO membranes as a function of applied voltage, we have performed *in-situ* XRD experiments. We used a Bruker D-8 Discover advanced XRD system (Cu Kα, $\lambda$ = 0.154 nm) to perform this experiment. A homemade XRD sample holder was designed to hold a few milliliters of water beneath the GO membrane, providing a continuous source of moisture to keep the membrane at 100% RH, mimicking the water permeation experiments. The interlayer spacing, *d*, was calculated using Bragg's equation, $d = \lambda/2sin\theta$, where $\theta$ is the scattering angle and $\lambda$ is the wavelength of the incident wave.

# Supplementary Information

1. **Fabrication of metal-GO-metal sandwich membranes and electrical measurements**

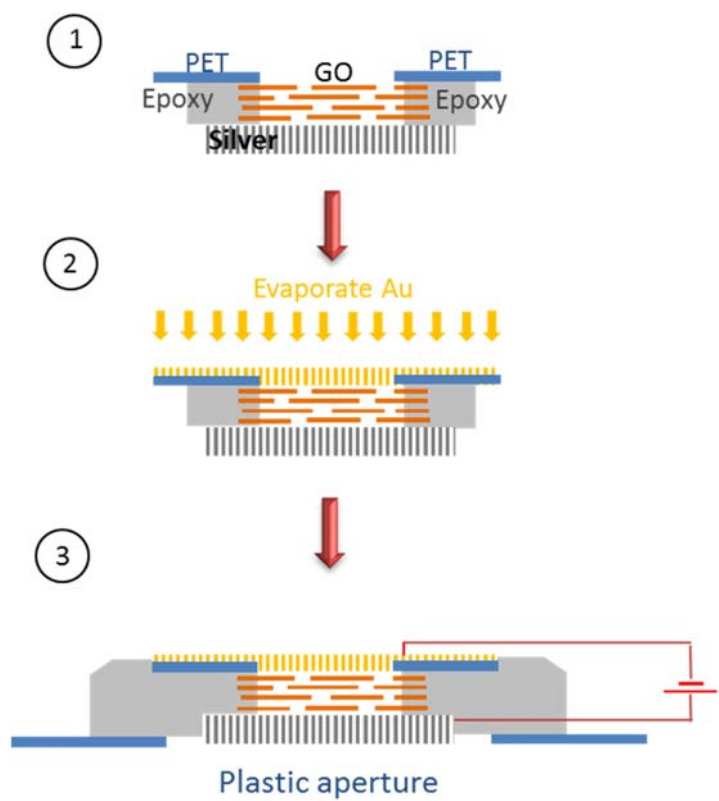

**Supplementary Fig. 1|** Fabrication procedures for the metal-GO-metal sandwich membrane.

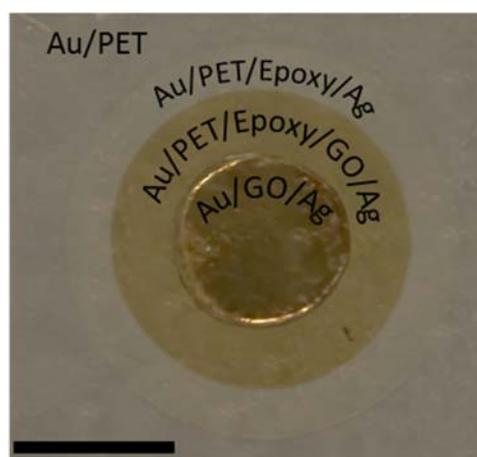

**Supplementary Fig. 2| Metal-GO-metal sandwich membrane device**. Photograph of one of our metal-GO-metal sandwich membranes attached to the PET sheet (step 2 in supplementary Fig. 1). Scale bar: 6 mm. This was further attached onto another plastic disc for sealing the metal container for gravimetric testing.

## 2. Conducting filament formation in GO membranes

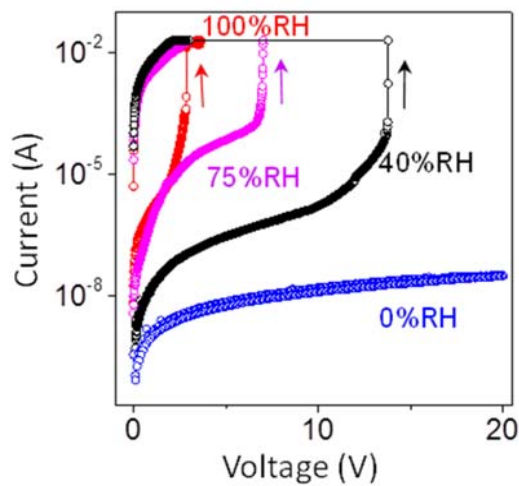

**Supplementary Fig. 3| Influence of humidity on conducting filament formation.** *I-V* characteristics during the first voltage sweep showing a sudden increase in the current for membranes exposed to humid conditions suggesting partial electrical breakdown of GO membrane and conducting filament formation.

It is known that in the presence of moisture on the surface of an insulator, the formation of permanent conducting paths, usually carbon, occurs across the surface of the insulator, known as tracking[1-3]. The tracking phenomenon is common in organic dielectrics and between layers of bakelite or similar dielectrics made of laminates. Consistent with this, we found that conducting filament formation in GO membranes is also highly sensitive to the humidity of the environment suggesting the occurrence of the tracking phenomenon. Supplementary Fig. 3 shows the *I-V* characteristics of the GO membranes at different relative humidity in the first voltage sweep. The sample exposed to zero humidity did not show any evidence of the formation of electrically conducting filaments even up to 50 V, confirming the high dielectric strength of GO in a dry atmosphere and consistent with previous reports[4,5]. However, the samples exposed to humid conditions deviate from this behavior, showing a sudden increase in the current at a certain voltage denoted as the critical voltage ($V_c$). After reaching $V_c$, the samples were permanently switched to a conducting state and their conductivity was found to be stable even after applying a negative voltage across the membrane (see main Fig. 1f). The decrease in value of $V_c$ with increasing relative humidity of the environment (supplementary Fig. 3) further suggests the contribution of absorbed water content (interlayer water) in the formation of conducting filaments inside the GO membrane, consistent with the tracking phenomenon.

To understand the electrical conductivity of GO membranes after the filament formation, we have measured the in-plane conductivity of the conducting GO membrane that displayed out-of-plane conductivity. For this, the membrane was peeled off from the silver substrate and then exfoliated using scotch tape. A pair of electrodes (3 mm apart with a width of 1 cm) was made on this freshly peeled GO membrane using conductive silver paste to measure the in-plane conductivity. Surprisingly, we found that the in-plane electrical conductivity of these GO membranes is similar to that of highly resistive pristine GO membranes (~15 µS/cm)[6]. Supplementary Fig. 4 compares the out-of-plane and in-plane *I-V* characteristics of the GO membrane after conducting filament formation.

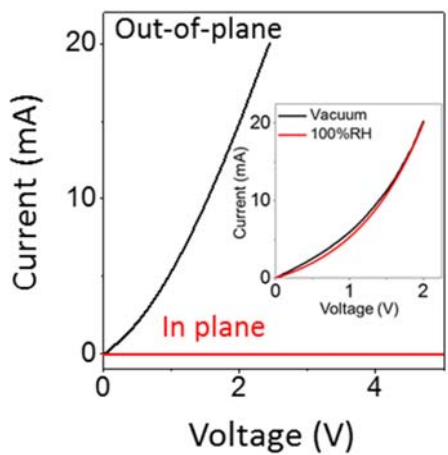

**Supplementary Fig. 4| In-plane and out-of-plane electrical conductivity.** In-plane and out-of-plane *I-V* characteristics of the GO membrane after the filament formation. Inset: Out-of-plane *I-V* characteristics of GO membrane at 100% relative humidity and vacuum.

To probe the stability of out-of-plane electrical conductivity after filament formation, we have further measured out-of-plane *I-V* characteristics of GO membrane at different humidity. Unlike the in-plane electrical conductivity in GO, which is sensitive to the humidity of the environment[7], the out-of-plane conductivity was found to be stable. The inset of supplementary Fig. 4 shows the out-of-plane *I-V* characteristics of the GO membrane at 100% relative humidity (RH) and vacuum. The stable out-of-plane electrical conduction in the GO membrane further agrees with the permanent filament formation in the GO membrane.

To demonstrate the existence of the conducting filaments inside the GO membrane, the top and bottom metal electrodes were peeled off using a scotch tape. The GO membranes were then further exfoliated using scotch tape to obtain the freshly produced inner surface for characterization using scanning electron microscopy (SEM), energy dispersive X-ray (EDX) spectroscopy, Raman spectroscopy, and X-ray photoelectron spectroscopy (XPS). The SEM image of a conducting GO membrane shown in supplementary Fig. 5a exhibits a texture similar to that of a pristine GO membrane with no noticeable features corresponding to the conducting filaments. EDX analysis further confirmed the presence of only carbon and oxygen elements in the membrane.

Raman spectra of the GO samples were collected using HORIBA's XploRA PLUS Raman spectrometer with an 1800 lines/mm grating and 532 nm laser excitation at a power of 1.35 mW and a spot size of 300 nm. Supplementary Fig. 5b and c show the Raman $I_D/I_G$ mapping for a pristine GO membrane and a GO membrane (close to the positive electrode) with conducting filaments (conducting GO membrane). The ratio $I_D/I_G$ is calculated from the intensity of the D band at 1351 cm$^{-1}$ to the intensity of the G band at 1594 cm$^{-1}$ in the Raman spectrum (supplementary Figs. 5e and f). As seen in supplementary Fig. 5b, the value of $I_D/I_G$ is uniform (0.93) over the whole area of 10 μm × 10 μm in the case of the pristine GO sample. However, for the GO membrane with conducting filaments, $I_D/I_G$ was found to be inhomogeneous and varied from 0.93 (blue-colored areas) to 1.1 (green-colored areas) at different spots on the sample. Typical Raman mapping showing the inhomogeneous $I_D/I_G$ ratio is shown in supplementary Fig. 5c. A similar increase in $I_D/I_G$ was reported for the reduction of graphene oxide and attributed to an increase in the $sp^2$ carbon network in the samples[8]. This suggests that the conducting filaments across the GO membrane are made up of $sp^2$ carbon.

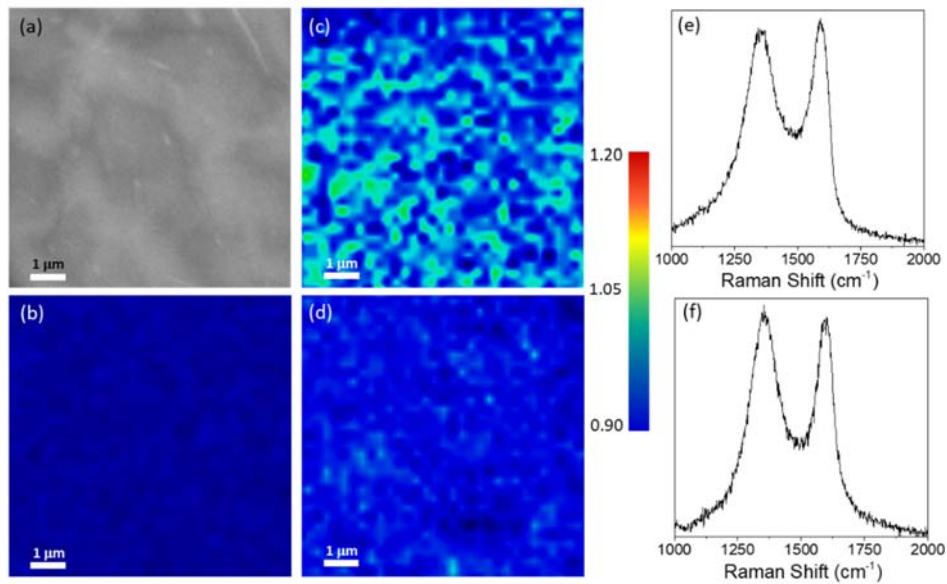

**Supplementary Fig. 5| Raman characterization of conducting filaments in a GO membrane:** (**a**) Topographical SEM image of a GO membrane after the formation of conducting filaments. Raman intensity ratio ($I_D/I_G$) mapping of D and G bands for (**b**) a pristine GO membrane and (**c** and **d**), a GO membrane after conducting filaments had been formed. (**c**) Raman imaging from the membrane surface close to the positive electrode (~ 200 nm away). (**d**) Raman imaging from the membrane surface close to the negative electrode (~100 nm away). (**e**) and (**f**) show the Raman spectra from the dark blue and green colored regions in (**c**), respectively.

To investigate the uniformity of carbon filaments across the top and bottom electrodes, we have carried out $I_D/I_G$ mapping experiments on a conducting GO membrane surface close to the negative electrode by multiple peelings of the membrane close to the positive electrode using scotch tape. The $I_D/I_G$ map obtained close to the negative electrode (supplementary Fig. 5d) shows the significant reduction in the size of domains with high $I_D/I_G$ ratio (~1.1) which clearly suggests the shrinkage of carbon filaments towards the negative electrode. This is further corroborated by XPS analysis where the $sp^2$ carbon content is found to be larger close to the positive electrode with no appreciable change observed in C/O ratio, with respect to pristine GO, either for the middle portion of the GO membrane or the membrane close to the negative electrode (see section 3). In addition, from the Raman mapping shown in supplementary Fig. 5c, it is apparent that the number density of conducting filaments is large and several filaments are formed in each GO flake (size of ~10 μm × 10 μm). The number of conducting filaments estimated by counting bright spots in supplementary Fig. 5d is ~$10^7$/cm$^2$. Based on this, the proposed conducting filament structure in GO membranes is shown in supplementary Fig. 6a.

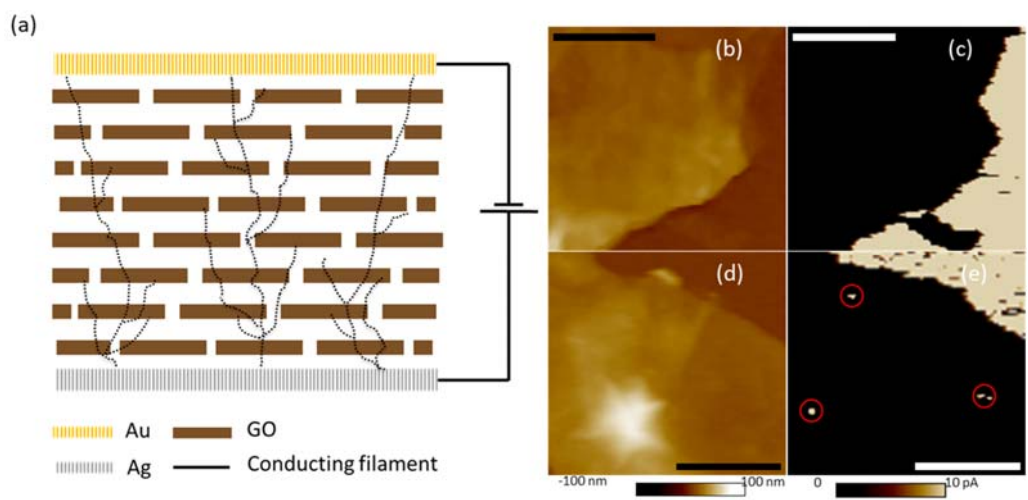

**Supplementary Fig. 6| Conducting carbon filaments in a GO membrane.** (**a**) Schematic showing the structure of conducting carbon filaments in the GO membrane. Topography and the corresponding TUNA current image of pristine GO (**b** and **c**) and conducting GO (**d** and **e**) membranes, respectively, exfoliated on a gold thin film coated Si substrate. Scale bars represent 1 µm.

Further, we have carried out atomic force microscope (AFM) imaging using Bruker Dimension ICON AFM operated in peak force tunnelling AFM (PF TUNA) mode to confirm filament formation in conducting GO membranes. PF TUNA imaging was carried out at 2.5 nN peak force using Bruker's PF TUNA probe (Au coating, spring constant of ~0.4 N/m). DC sample bias was varied from 3 to 5 V and the gain setting was changed from $10^9$ V/A to $10^{10}$ V/A for different samples. In particular, the surface close to the negative electrode of conducting GO membranes was exfoliated to validate the presence of filaments and pristine GO membrane was used as a reference. Both the pristine and conducting GO membranes were exfoliated on Cr/Au thin film (5 nm/95 nm) deposited on a Si substrate. This set-up facilitates the application of a constant DC bias between the sample and the AFM probe in PF TUNA mode. The current passing through the sample between the Cr/Au thin film and the AFM probe is measured using a current sensor when the probe and sample are intermittently brought into contact in each tapping cycle. Thus, the mapping of electrical current across the samples provides their TUNA current image along with the topography.

Supplementary Fig. 6b and d show the height images of pristine and conducting GO membranes exfoliated on the Cr/Au deposited Si substrate. Their corresponding TUNA current images are shown in supplementary Fig. 6c and e. Topography of both the membranes in the height images did not show any significant difference and the measured thickness of membranes was varied between ~30 to 40 nm. However, the TUNA current images of pristine and conducting GO membranes exhibited considerable changes. As expected the TUNA current image of pristine GO membrane shows an apparent contrast for GO (dark regions) from the surrounding gold thin film because the magnitude of TUNA current between the gold thin film and the probe is much higher than that between GO and the probe (due to the insulating behaviour of GO). Therefore, the gold thin film region appears brighter than the GO region. On the other hand, the TUNA current images (supplementary Fig. 6e and Fig. 1 in main text) recorded for conducting GO membranes reveal the presence of small conducting regions (which cannot be identified in height images) within the non-conducting GO region (dark region). We have further carried out PF TUNA imaging of several samples of both pristine GO and conducting GO membranes and found the presence of small conducting regions only in the conducting GO membrane samples. Further, we have estimated the filament density to be ~$10^7$/cm$^2$ by counting the conducting regions in the TUNA current image, which is in agreement with the density obtained from the Raman analysis.

As shown in the schematic, these conducting filaments across the top and bottom electrodes mimic several resistors in parallel and accordingly the estimated current through each filament is ~1 nA for a 1 V applied potential. To validate the parallel resistor model, we have divided a large conducting GO membrane (7 mm diameter) into four equal pieces. We found that the resistance of the four individual pieces is relatively similar in magnitude and is four times larger than that of the large parent membrane (supplementary Fig. 7) but with a similar resistivity. This suggests that the high resistance of small pieces could be due to the lower number of parallel filaments which further confirms the parallel resistor model. In addition, this experiment also confirms the uniform distribution of the filaments inside the membrane.

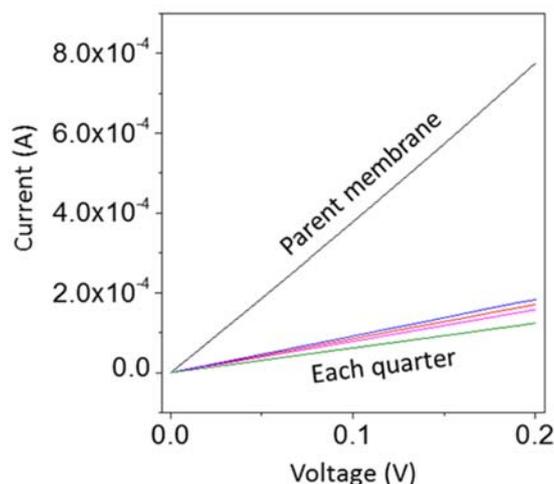

**Supplementary Fig. 7| *I-V* characteristics of GO membrane with conducting filaments.** Out-of-plane *I-V* characteristics of a conducting GO membrane with a diameter of ~7 mm before (parent membrane) and after dividing into four equal pieces.

3. <u>**X-ray photoelectron spectroscopy (XPS) on GO membranes**</u>

To investigate the chemical stoichiometry of GO membranes before and after the application of an electric potential (after electrically controlled water permeation experiments), we performed XPS. XPS spectra were acquired in an ultrahigh vacuum system with a base pressure of $< 3 \times 10^{-10}$ mbar using a monochromated Al Kα source at 1486.6 eV (Omicron XM 1000) and a power of 220 W. An aperture diameter of 2 mm was used with the sample normal at 45° to both the X-ray source and the entrance optics of the hemispherical energy analyser (Omicron EA 125).

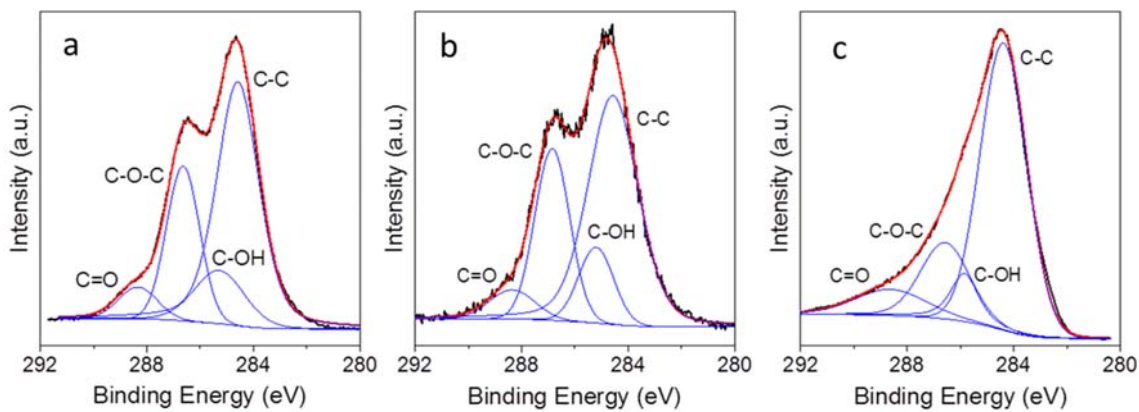

**Supplementary Fig. 8| X-ray photoelectron spectroscopy (XPS) characterization of GO membranes.** (**a**) XPS spectrum from a pristine GO membrane. (**b** and **c**) XPS spectra from GO membranes used for the electrically controlled permeation experiments. (**b**) shows a spectrum from a freshly cleaved membrane surface close to the inner middle region and (**c**) is from a freshly cleaved surface close to the positive electrode. Black lines: raw data; red lines: the fitting envelope; blue lines: deconvolved peaks attributed to the chemical environments indicated.

Supplementary Fig. 8a shows the XPS spectra from the pristine GO membrane and supplementary Fig. 8b and c represent XPS spectra of a GO membrane after electrically controlled permeation acquired from an inner surface of the membrane and a surface close to the positive electrode, respectively. A membrane surface from close to the electrode was obtained by removing the silver electrode through mechanical peeling, and then cleaving the membrane using scotch tape to reveal an inner surface. Using XPS Peak 4.1, each C 1$s$ spectrum was fitted with four components representing the main bonding environments found in graphene oxide: C-C (284.5-284.8 eV), C-OH (285.2-285.4 eV), C-O-C (286.3-286.9 eV), and C=O and C(=O)-(OH) (287.8-289.1 eV)[9,10]. C/O ratios calculated from the fitted peak areas were found to be similar (3.2) for pristine GO and the inner surface of the membrane after electrically controlled permeation. In contrast, the membrane surface close to the positive electrode shows an increase in C/O ratio (3.6) indicating a higher $sp^2$ fraction close to the electrode. C-C fractions are found to be 56% and 63% respectively for the inner surface and from the surface close to the electrode. This increase in $sp^2$ fraction close to the positive electrode is attributed to the formation of conducting carbon filaments after the application of a voltage across the GO membrane where the concentration of filaments is expected to be large close to the positive electrode (supplementary Fig. 6a).

4. **Mass spectrometry**

Electrical control of water permeation through GO membranes was also probed using mass spectrometry (MS), schematically shown in supplementary Fig. 9a. Here, a Au/GO/Ag sandwich structure was placed between two rubber O-rings in a custom made permeation cell (supplementary Fig. 9a). Copper leads from the top and bottom electrodes were connected to a sourcemeter via an electrical feedthrough. The permeate side pressure was maintained at $10^{-6}$ bar. Water vapour (25 mbar) and helium (25 mbar) were fed into the top chamber and permeation through the sample was monitored at the permeate side using MS. We used a quadrupole residual gas analyzer (HPR 30 Hiden Analytical) to measure the partial pressure of permeated species and wet cotton in the top chamber as a constant feed for water vapour. Supplementary Fig. 9b shows the partial pressure of water ($P_{H2O}$), hydrogen, oxygen, and helium (He) in the permeate side at different current through the membrane. No appreciable change in helium partial pressure is observed during voltage cycling, indicating that the conducting GO membranes are impermeable to helium, as reported before[11]. This also confirms that the membranes are not damaged by the application of electric fields. Supplementary Figs. 9b and c show that $P_{H_2O}$ decreases with increasing current through the membrane suggesting a decrease of water permeation through the GO membrane with increasing current. This is consistent with the gravimetric measurements reported in the main text. Additionally, the lack of any significant change in $H_2$ or $O_2$ partial pressure at the permeate side further indicates that $H_2$ and $O_2$ are not released during voltage cycling, ruling out the possibility of electrolysis of water. Supplementary Fig. 9b also confirms the reversible control of water permeation by electrical means.

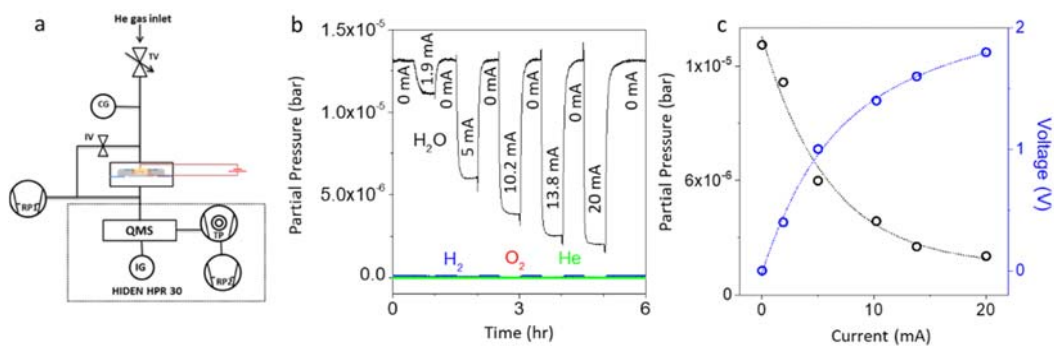

**Supplementary Fig. 9| Mass spectrometry for probing electrically controlled water permeation.** (**a**) Schematic of the experimental setup for mass spectrometry (MS) measurements. A throttle valve (TV) controls the gas inlet with a capacitance gauge (CG) used to measure the upstream pressure. An isolation valve (IV) isolates upstream and downstream sides of the membrane. A rotary pump (RP1) evacuates the feed as well as the permeate side to 1 mbar. The quadrupole mass spectrometer (QMS) measures the downstream partial pressure. A turbomolecular pump (TP) backed by a rotary pump (RP2) evacuates the high vacuum chamber of the mass spectrometer. An active ion gauge (IG) measures the pressure down to $1 \times 10^{-9}$ Torr in the high vacuum side. (**b**) The partial pressure of He, $H_2$, $O_2$ and $H_2O$ at the permeate side as a function of time at different current through the membrane. No detectable change is observed in the partial pressure values of He, $H_2$ and $O_2$ under different current through the membrane. (**c**) The partial pressure of $H_2O$ as a function of the current across the GO membrane, and the corresponding *I-V* characteristics (color-coded axis). The dotted lines are guide to the eye.

5. **Current controlled permeation**

To understand the influence of both voltage and current, we performed electrically controlled water permeation through GO membranes with different permeation areas and thicknesses. Supplementary Fig. 10a shows results for permeation through 1 μm thick GO membranes with different permeation areas as a function of electric current. The current required to achieve similar values of permeation rate was found to be four times smaller for a membrane with an area that is four times smaller. This suggests that current density controls water permeation through the membrane. The influence of current on the permeation can be further seen in the experiment performed on membranes with different thicknesses. Supplementary Fig. 10b shows the normalized water permeation rate through 1 μm and 5 μm thick GO membranes as a function of current. As seen in the figure, the rate of decrease of water permeation rate for both 1 and 5 μm samples is similar, however, the voltages required to maintain the same current are significantly different. The direct correlation between the current and the permeation rate further confirms that water permeation is mainly controlled by the current and not the voltage.

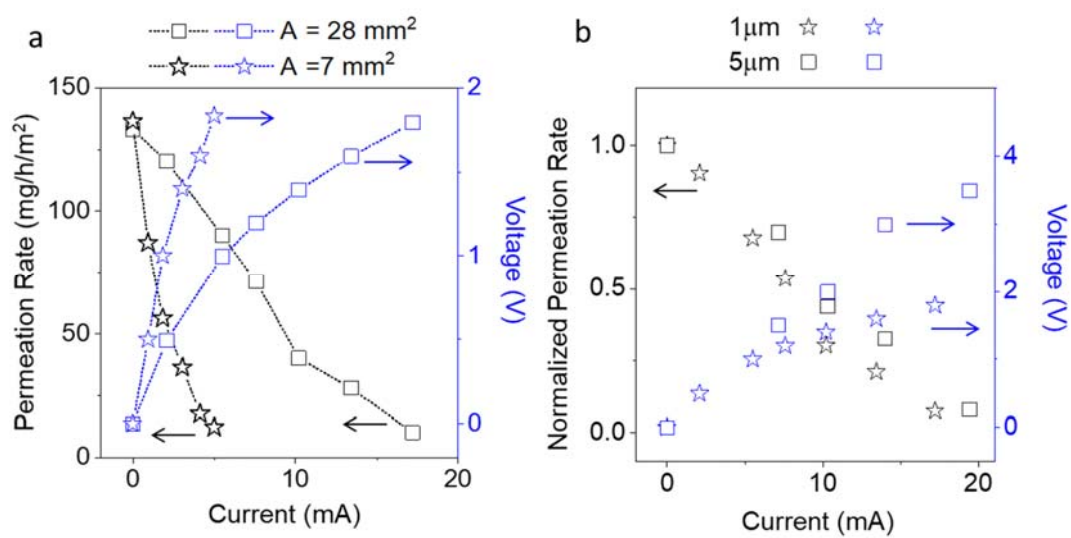

**Supplementary Fig. 10| Current controlled permeation**. (**a**) Permeation rate as a function of the current through two different GO membranes with different areas (28 mm$^2$ and 7 mm$^2$) and the corresponding *I-V* characteristics (color-coded axis). (**b**) Normalized water permeation rate as a function of the current through GO membranes with two different thicknesses and the corresponding *I-V* characteristics (color-coded axis). Permeation rates were normalized with respect to zero applied voltage because the absolute water permeation rates of 1 μm and 5 μm GO membranes were different.

6. **Joule heating effect**

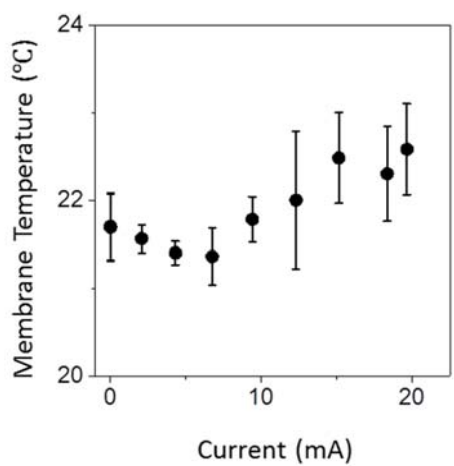

**Supplementary Fig. 11| Joule heating in a GO membrane.** Measured membrane temperature as a function of the current flowing across the membrane.

To probe electric-current-induced joule heating, we used an infrared thermometer (N92FX, Mapline Inc.) to monitor the variation of membrane temperature as a function of electric current across the membrane. Several measurements were performed at more than 10 different locations to obtain the average temperature for each current with the results shown in supplementary Fig. 11. No significant changes in the membrane temperature were found. A temperature increase of only ≈ 1 °C was measured for a current of 20 mA which eliminates joule heating effects in our water permeation experiments.

7. **Electric field due to a current-carrying conductor**

To explain the electrically controlled water permeation through GO membranes, we propose a simple model. We consider the case of a single conductive filament of length $L$ and radius $a$ ($a << L$) carrying a constant stationary current of $I$ in a closed circuit with an applied potential of $V$. It is known that current-carrying conductors produce an electric field $E(r,z)$ associated with the electric potential $\psi(r,z)$ around the conductor, depending on its dimensions and conductivity ($\sigma$)[12-14]. One can envisage that this potential $\psi(r,z)$ subsequently decays to zero at a point (at distance $b$) far from the filament. For any point at a distance $r$ between $a$ and $b$, Laplace's equation is valid[15]:

$$\nabla^2 \psi(r,z) = 0 \qquad (1)$$

The boundary conditions for the above scenario are $\psi(r,z) = -Jz/\sigma$ and $\partial \psi/\partial z = -E_0 = -J/\sigma$ when $r \leq a$ and $\psi(r,z) = 0$ when $r = b$. Here, $J$ is the current density through the filament, $\sigma$ is the conductivity of the wire, and $z$ varies from 0 to $L$ (supplementary Fig. 12a).

Solving Eq. (1) under the above boundary conditions[14] in cylindrical coordinates (r,ϕ,z) we obtain,

$$\psi(r,z) = \frac{Jz}{\sigma} \frac{\ln(r/b)}{\ln(b/a)} \qquad (2)$$

The electric field associated with $\psi(r,z)$ is calculated as

$$E(r,z) = -\left(\frac{\partial \psi}{\partial r}\hat{r} + \frac{\partial \psi}{\partial z}\hat{z}\right) = \frac{Jz}{\sigma r}\frac{1}{\ln(a/b)}\hat{r} + \frac{J}{\sigma}\frac{\ln(r/b)}{\ln(a/b)}\hat{z} \qquad (3)$$

At very close distances from the filament, the magnitude of the radial component of electric field is higher than that of the z component. In our experiments, ~1 nA flows through a single conductive filament of $L = 1$ μm for an applied potential of 1 V. All these filaments are surrounded by the dielectric GO regions as confirmed from the Raman mapping and AFM experiments. Hence, the electric potential $\psi(r,z)$ decays to zero at some arbitrary radial distance $b$ from the filament. Considering $b = 500$ nm (since the average separation between conductive filaments is ~1 μm, as determined by the Raman measurements) and assuming $a = 10$ nm (from PF TUNA imaging), the magnitude of the electric field and its spatial distribution is plotted as a function of $r$ and $z$ in supplementary Fig. 12b. As seen in the figure, we found that the field remains high close to the filament (< 10 nm) and it decays slightly (by ~6 times) from the top positive electrode (2.3 × 10$^7$ V/m) to a point at distance of 100 nm above the bottom electrode (4 × 10$^6$ V/m). It is noteworthy that the field around the filament persists even up to a radius of 50 nm with a magnitude varying from 4.3 × 10$^6$ V/m to 5.4 × 10$^5$ V/m depending on the distance from the top positive electrode. We have also found that for the same 1 nA current, the purely radial component of $E$ decreases by

~40 times as the radius of the filament is increased from 10 nm to 100 nm. In summary, the current carrying filaments in GO membranes produce a radial electric field that is sufficiently strong to dissociate water molecules into hydroxide (OH⁻) and hydronium ($H_3O^+$) ions. Importantly, all of the above estimations are based on a simple model of single straight conducting wire (zero order approximation). However, the complicated structure of conducting filaments (as shown in supplementary Fig. 6) could produce even higher electric fields due to the close arrangement of individual filaments, especially near the positive electrode.

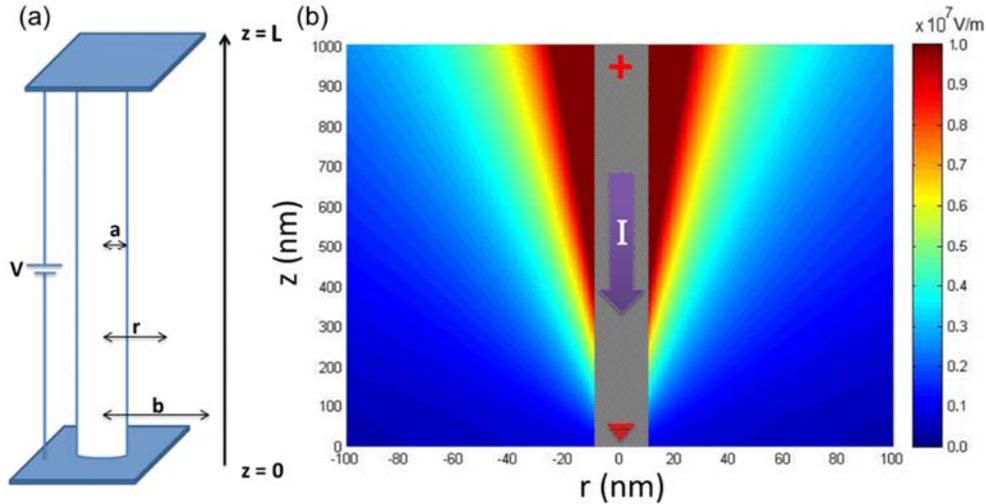

**Supplementary Fig. 12| Electric field around a current carrying conductor.** (**a**) Schematic showing the application of a voltage $V$ across an electrically conducting wire with radius $a$ and length $L$. $b$ is a point at which the potential decays to zero. $r$ represents any point between $a$ and $b$ where the electric field, $E$, is calculated. (**b**) Magnitude of $E$ and its spatial distribution as a function of $r$ and $z$ around a conductive filament with 1 V potential difference across the ends and with 1 nA current flow.

8. **Electric-field-induced dissociation of water and its influence on water permeation**

Random fluctuations in molecular motion are known to occasionally produce an electric field strong enough to dissociate water (known as autoionization), into hydroxide (OH⁻) and hydronium ($H_3O^+$) ions[16]. Ions produced in this manner recombine very quickly because molecular fluctuations vanish within tens of picoseconds. It is also known that a high external electric field is sufficient to dissociate water[17-19]. Electric-field-enhanced dissociation of water (without producing $O_2$ and $H_2$ gas) has been observed previously when an electric current is driven through bipolar membranes[18]. Based on this, we propose that the strong electric field near to the conducting filaments in the GO membrane dissociates water in the interlayer channels of the membranes into OH⁻ and $H_3O^+$ ions according to eq. (4).

$$(n + m + 2)H_2O \leftrightarrows OH^-\cdot n(H_2O) + H_3O^+\cdot m(H_2O) \qquad (4)$$

By increasing the electric current in the conducting filaments, the electric field (eq. 3) increases and hence the water dissociation rate increases. Once the electric current is switched off, the field vanishes and hence the reaction towards the left side in eq. 4 becomes more favorable.

The rate of reaction for the ionization in eq. 4 depends on the activation energy. In bulk water the activation energy originates from the above mentioned random fluctuations in the molecular motions[16] which causes an electric field strong enough to break an oxygen–hydrogen bond, resulting in hydroxide (OH⁻) and a hydronium ion ($H_3O^+$). However, it is known that the effective activation energy decreases in the presence of an electric field[19] due to the induced local dipoles in the system, supporting the proposed model.

To understand the influence of dissociation of water on the water flow through the graphene capillaries in the GO membranes we performed non-equilibrium molecular dynamic simulations (MD simulations). By using the large scale atomic/molecular massively parallel simulator LAMMPS[20], we investigated the dynamical properties (i.e. flow rate) of a mixture of hydronium, hydroxide, and water inside a graphene capillary by using a previously employed model[11], see supplementary Fig. 13a.

For the MD simulations the rigid model was used for hydronium ($H_3O^+$) and hydroxide (OH⁻) as presented in previous studies[21,22]. The graphene layers were kept fixed, and the SPC/E model was employed to describe the water molecules. The carbon and oxygen atoms interact via Lennard-Jones (LJ) pair potentials ($\sigma_C$ = 0.0553 kcal mol⁻¹, $\varepsilon_C$ = 3.4 Å, $\sigma_{H3O+}$ = 0.147467 kcal mol⁻¹, $\varepsilon_{H3O+}$ = 3.05 Å, $\sigma_{OH-}$ = 0.149618 kcal mol⁻¹, and $\varepsilon_{OH-}$ = 3.84 Å) and cross LJ potential parameters were obtained by the Lorentz-Berthelot combining rules. The cut-off radius for the LJ potential was chosen at 10 Å. The NVT ensemble (Nos'e-Hoover thermostat) was used to control the temperature for both non-rigid and rigid molecules at room temperature. A particle-particle particle-mesh (pppm) was used to compute the long-range Coulomb interaction with a desired relative error in the forces for long-range Coulomb interactions solvers of 10⁻⁴. In all cases, the time step was chosen as 1 fs. To validate the force fields used, we have calculated the diffusion coefficient of bulk water and found that it is in good agreement with previous experimental results, i.e., $D_0$ = 2.45 × 10⁻⁵ cm²/s.[23]

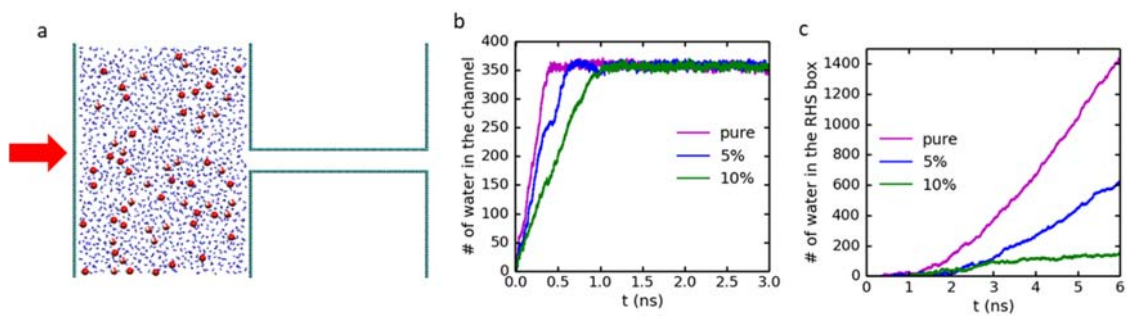

**Supplementary Fig. 13| Molecular dynamic simulations.** (**a**) Side view of our MD simulation setup used to study the flow of water mixed with hydronium and hydroxide ions in the graphene capillary. The model contains two boxes which are connected by a graphene capillary. At the beginning of the simulation, water was mixed with hydronium and hydroxide (red and white dots) ions. By moving the left wall (subjected to external pressure) of the box towards the capillary, the water flow is created and the right box is gradually filled. The arrow indicates the direction of the applied external pressure on the left wall of the box. (**b**) Number of water molecules in the capillary and (**c**) number of water molecules in the right box for pure water and water with ions once pressure is applied to the left box.

Supplementary Fig. 13a shows the simulation box, which contains two compartments that are connected by a graphene capillary of height 10 Å. Periodic boundary conditions were applied along the z-direction in the boxes and along the y-direction in the capillary. The simulation unit cell contained 4362 water/ion molecules, and each graphene capillary has a size of 8 × 2 nm$^2$ (720 carbon atoms). First, the molecules inside the left side box (supplementary Fig. 13a) were relaxed in order to reach its equilibrium for 1 ns. Afterward, we applied a pressure of 1 bar on the vertical wall of the left box (as shown by arrows in supplementary Fig. 13a) to move the wall toward the capillary so that the water/ion molecules enter the capillary and flow toward the right box. The rate of filling of the capillary and the right box was found to depend significantly on the number of H$_3$O$^+$ and OH$^-$ in the system (supplementary Figs. 13b and c). We found that increasing the concentration of ions, decreases the water flow rate through the capillary. When the ion concentration reaches ~10% (by number) the water flow rate significantly reduced (supplementary Fig. 13c). It is noteworthy that the 10% value is the concentration of ions in the left box however the concentration of ions in the capillary is only 2%, which indicates a small fraction of ions is required to reduce the water permeation significantly.

The observed decrease in water flow rate with increasing ion concentration in the capillary could be due to ion hydration effects; with an increasing number of ions in the graphene capillary, water tends to remain in the capillary thereby decreasing the water flow. It is also worth mentioning that the concentration of ions found to reduce the water flow rate in the MD simulation could be an overestimated value. The free space available in the interlayer channels of GO is < 1 nm, as used for the MD simulations and the functional groups inside the GO capillary may absorb ions and making the channel more hydrophilic. These effects were not included in the MD simulations. Nevertheless, our MD simulations suggest that the dissociated water molecules inside the interlayer channels of GO membrane could significantly affect water permeation through the membrane. It is also interesting to note that a high concentration of dissociated water (~50%) was also observed in interfaces and metallic surfaces and shows that they could be energetically stable even at room-temperature[24].